\documentclass{desyproc}
\usepackage{subfigure}
\begin{document}
\vspace{-1.0in}
\title{Exclusive High Mass Di-leptons in CDF}

\author{{\slshape Michael Albrow}\\[1ex]
Fermi National Accelerator Laboratory, Batavia, IL60510, USA\\}

\maketitle

\begin{abstract}
Talk given at 13th International Conference on
Elastic and Diffractive Scattering (13th Blois Workshop), CERN June 2009

\vspace{2mm}
In the Collider Detector at Fermilab, CDF,
we have measured central exclusive production, $p+\bar{p}\rightarrow p+X+\bar{p}$, where $X$ is a pair
of leptons or photons and nothing else. In this talk I focus on
central masses $M(X) >$ 8 GeV/c$^2$. We measured QED production $\gamma\gamma \rightarrow e^+e^-,\mu^+\mu^-$ up
to $M(X)$ = 75 GeV/c$^2$, and candidates for photoproduction of Upsilons, $\gamma I\!\!P \rightarrow
Y(1S),Y(2S),Y(3S)$. I report a search for exclusive photoproduction of $Z$-bosons, and the status of searches
for exclusive two-photons: $p+\bar{p}\rightarrow p+\gamma\gamma+\bar{p}$. These measurements constrain the cross section
$\sigma(p+p\rightarrow p+H+p)$ at the LHC.

\end{abstract}

\section{Introduction}

\hspace{.5in}By central exclusive production CEP (also called central exclusive diffraction, CED) at the
Fermilab Tevatron we mean reactions $p+\bar{p}\rightarrow p+X+\bar{p}$, where $X$ is a simple system fully measured,
and ``+" are large rapidity gaps ($\Delta y \gtrsim 5$ units) with \emph{no} particles. In the Collider
Detector at Fermilab, CDF~\cite{cdf}, we cannot detect the forward $p$ or, except for some large $M(X)$ events, the
 $\bar{p}$. However we installed scintillation counters (beam shower counters, BSC) along the beam pipe, which
 detected showers from particles out to pseudorapidity $|\eta|$ = 7.4. Requiring them to be empty
 selects events in which the $p$ and $\bar{p}$ did not fragment and went down the beam pipe.
 
 In the Standard Model the only significant coherent $t$-channel exchanges over such large rapidity gaps are
 color singlets with charge $Q=0$ and spin $J$, or effective spin $\alpha(t=0)$, $\geq$ 1. These are the photon,
 $\gamma$, pomeron $I\!\!P$ (C=+1) and odderon $O$ (C = -1). $Z$-boson exchange would be allowed, but the proton
 would inevitably break up. The odderon has not yet been convincingly observed; in our observation of exclusive
 $J/\psi$ and $\psi(2S)$ reported in this meeting by Pinfold~\cite{cdfchic,jimp} we placed a new limit. In this talk I will report
 on CDF measurements of exclusive lepton pairs, $p+\bar{p} \rightarrow p + l^+l^- + \bar{p}$ above the charmonium
 region, see Fig.1. This includes non-resonant (QED) $\gamma\gamma\rightarrow e^+e^-,\mu^+\mu^-$, and photoproduction : $\gamma I\!\!P
 \rightarrow \Upsilon, Z$. We can also search for exclusive $I\!\!P I\!\!P \rightarrow 
 \chi_b \rightarrow \Upsilon+\gamma$, but the cross section is very small and its observation probably requires no 
 additional collisions (no pile-up). This
 process, like exclusive $\chi_c$, is a good test of exclusive Higgs boson production, as the QCD part of the Feynman
 diagrams is identical. While our CDF studies provide good tests of QCD with large rapidity gaps, and hard pomeron behavior,
 they are also precursors to processes with $M(X) \gtrsim $ 100 GeV/c$^2$ at the LHC~\cite{fp420}, where $X$ can be 
 $h, H, W^+W^-, \tilde{l}^+\tilde{l}^-$ etc., including 
 any exotic particles that couple to gluons or photons and have the right quantum numbers. If states such as Higgs bosons
 are seen this way at the LHC, their mass, width, spin, C-parity and coupling $\Gamma(Xgg)$ can be determined in unique ways.
 Even a pair of nearby states, e.g. MSSM $h(140)\rightarrow b\bar{b}$ and $H(150)\rightarrow b\bar{b}$ can be resolved, which
 is impossible by other means.
 
 \section{CDF detectors}
 
 The central detector of CDF~\cite{cdf} has layers of silicon tracking and drift chambers surrounded by a time-of-flight scintillation counter
 barrel,
 in a solenoidal field. This is surrounded by electromagnetic and hadronic calorimeters, muon scintillators and 
 tracking chambers. The forward region, $\theta < 3^\circ$, has, on each side, a 48-channel Cherenkov luminosity counter
 hodoscope, a ``miniplug'' calorimeter, a set of beam shower counters, BSC, and on the outgoing $\bar{p}$ side, scintillating
 fiber trackers in Roman pots. The BSC were very important as rapidity gap detectors in no-pile-up events, and for triggering on exclusive events. 
  They are relatively simple scintillation counters around the beam pipes covering $ 5.5 < |\eta| < 7.4$. 
  Only BSC-1 sees primary particles, and it has two radiation lengths in front to convert photons; the others see showers created in the
 beam pipe. We have proposed them for CMS~\cite{fsccms}; all LHC experiments should have them! Another recommendation to all
 experiments is to record zero-bias, or bunch-crossing, triggers routinely, e.g. at 1 Hz. These were essential in our CDF
 exclusive studies. We divided those events into two classes: [a] = probably no interaction, e.g. no tracks, and [b] at least one
 interaction, with tracks from the beam line. Then for each subdetector, e.g. BSC-1 which had 8 PMTs, we plotted the ``hottest''
 PMT as Log(max ADC counts in BSC-1) for events in classes [a] and [b]. One can also plot the sum $\sum_i(ADC_i)$. Repeating for all
 subdetectors, one finds cuts that select events with all the CDF detectors empty, except for the state $X$. 
 
 \begin{figure}[htp]
\centering
\subfigure[Two-photon production of lepton pairs]{
  \label{fig:qed}
  \includegraphics[width=.35\textwidth]{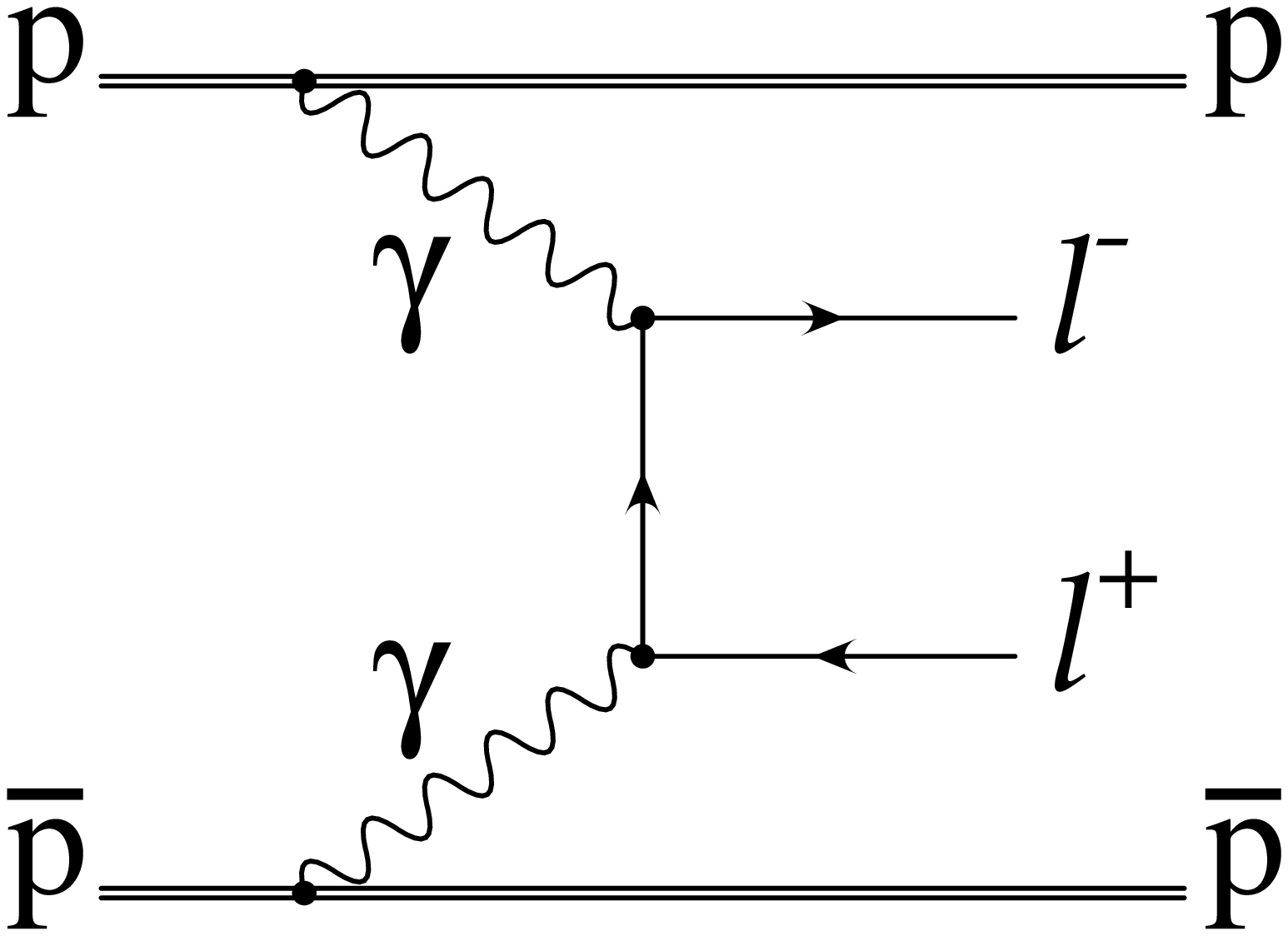}}
  \hspace{.3in}
  \subfigure[Z photoproduction (or $\Upsilon$ photoproduction if $q = b$).]{
  \label{fig:zphot}
  \includegraphics[width=.35\textwidth]{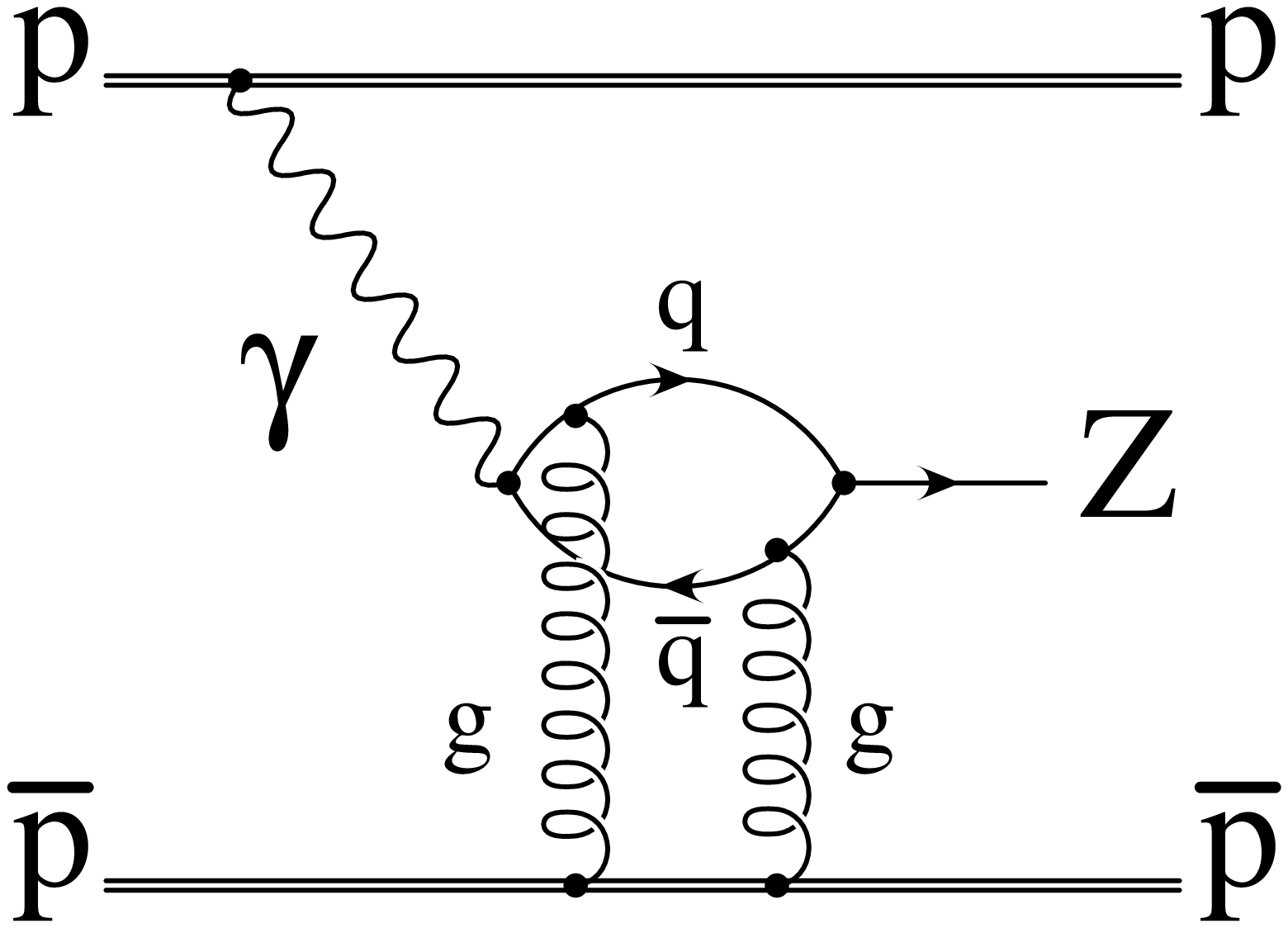}}
  \caption{Exclusive two-photon production of lepton pairs, and exclusive photoproduction.}
  \end{figure}
  
 \section{CDF exclusive physics program}
 
 In March 2001 some of us proposed~\cite{cdfloi} to add very forward proton tracking detectors to CDF to look for exclusive Higgs boson
 production. At that time some theorists (cited in Ref.~\cite{cdfloi}) had suggested that the cross section could be as 
 high as 10-100 fb, and a signal could be seen using the missing mass technique~\cite{ar}. The Durham group 
 prediction~\cite{kmrcdfh} was a factor $\gtrsim 10^2$ lower, at 0.06 fb (for M(H) = 120 GeV), impossibly low for the Tevatron. 
 Incidentally in Ref.~\cite{cdfloi} we suggested for the first
 time that exclusive $\gamma\gamma$ production is a good test of the theory, and that troublesome pile-up can be highly
 suppressed with ``fast timing Cherenkov counters". Given the very large theoretical uncertainty on $\sigma(H)$, the proposal could not
 proceed at that time, but we embarked on a program to measure related exclusive processes with cross sections accessible
 even in the Durham model. We have now measured $p+X+\bar{p}$ final states with $X = \gamma\gamma, \chi_c$ and dijets
 $JJ$, all of which have very similar QCD diagrams and issues, and all three are consistent with the Durham group
 predictions (which have a factor ``a few'' uncertainty). One can now be certain that it is possible to produce a
 Higgs boson (if it exists) at the LHC with no other particles, and reasonably confident that their prediction
 $\sigma(SMH(120)) \sim$ 1 - 10 fb at $\sqrt{s}$ = 14 TeV is not unrealistic. The cross section can be much higher in some Beyond
 SM scenarios. 
 
 The exclusive processes measured in CDF have different strengths and weaknesses. $X = \gamma\gamma$ is the cleanest as
 the photons, like the Higgs, have no strong interactions. Looking at the Feynman diagrams through a ``QCD-only filter" they are
 identical; simply the $q$-loop (mostly $u$ and $c$) becomes a $t$-loop and $\gamma\gamma$ is replaced by $H$. But the cross
 section is very small. The Durham prediction~\cite{kmrgg} is 36 fb for $|\eta(\gamma)| < 1$ and $E_T(\gamma) > 5$ GeV.
 This corresponds to 0.8$^{+1.6}_{-0.5}$ events in the CDF search~\cite{cdfgg}; 3 candidates were found, of which 2 were
 ``perfect" with two single narrow electromagnetic showers, while the third had some characteristics of 
 $\pi^0\pi^0$ (broader showers). Note
 that the two golden candidates were distilled from $10^{12}$ inelastic collisions; such things can be done! $X=\chi_c$ is
 also clean and has a much bigger cross section: $\frac{d\sigma}{dy} (y=0) = 76 \pm 10(stat) \pm 10(syst)$~nb
 (CDF~\cite{cdfchic,jimp}), compared with 90 nb (with a large uncertainty) predicted by Durham~\cite{kmrchic}. The weakness
 is that the charm mass is small so the process is not very perturbative, the $\chi_c$ is colorless but it is still a hadron
 with final state interactions, and resolving $\chi_{c0}(1P), \chi_{c1}(1P),$ and $\chi_{c2}(1P)$ is difficult. The
 $\chi_b$ is more perturbative than the $\chi_c$, so the theory is under better control, but the cross section is expected 
 to be only about 1/500th that
 of the $\chi_c$ and the decay modes are not well known. 
 For the $\chi_{b0}$ the Particle Data Group gives only $\chi_{b0}(1P) \rightarrow \gamma\Upsilon(1S) <$ 6\%, with no other modes well known, so it will not be possible to give a cross section soon.
 (Some hadronic decays with poorly known branching fractions are impossible to trigger on in a normal collider environment.)
 Finally CDF also measured~\cite{cdfjj} exclusive di-jets, produced by the process $gg \rightarrow gg$ with a color-cancelling gluon
 exchange. The cross section is quite large, $\approx$ 100 pb for two jets with $E_T > 15$ GeV and $|\eta| < 2.5$, with
 the dijet having $>$ 80\% of the total central mass, i.e. $R_{JJ} > 0.8$. In this region there is an excess of dijets
 compared with inclusive dijet expectations. The data, out to Jet $E_T^{min}$ = 35 GeV, agree within a factor $\sim 3$
 with the Durham prediction~\cite{durhamjj}, but are more than an order of magnitude lower than the \textsc{dpemc}
 prediction~\cite{dpemc}. 
 
 The main particle states that can be produced exclusively in double pomeron exchange, $D I\!\!P E$, are
 $\sigma(600),f_0(980), \chi_c, \chi_b$ and $H$. The first two were observed at the ISR~\cite{akes} and, less cleanly, at the SPS
 (fixed target), and the $\chi_c$ is now established~\cite{cdfchic,jimp}. The other accessible states are $\gamma\gamma$ and $JJ$.
 It will be very hard to get a good ($\lesssim$ 25\%) measurement of the $\chi_b$, but $H$ appears to be in reach at the LHC, provided
 the two protons can be measured. (Double proton tagging for $p+\chi_b + p$ at high enough luminosity to get some events is
 probably not feasible, due to its low mass.)

  In CDF we observed $\gamma\gamma \rightarrow e^+e^-$ collisions with $E_T(e) >$ 5 GeV for the first time in hadron-
  hadron collisions. We published~\cite{cdfee} 16 events (backgound $\sim$ 1.9) with $M(e^+e^-) > $ 10 GeV/c$^2$ in excellent agreement
  with the \textsc{lpair} Monte Carlo. The highest mass event was at 38 GeV/c$^2$, and all pairs are very back-to-back with
  ($180^\circ - \Delta\phi) <
  2.4^\circ$. We followed that with a search~\cite{cdfz} for exclusive $Z$, in the process finding eight $e^+e^-$ or
  $\mu^+\mu^-$ events with $M(l^+l^-)$ from 40 to 75 GeV/c$^2$. Again ``QED rules", and the lepton pairs 
  all have $(180^\circ - \Delta\phi) < 0.75^\circ$. This mass reach is as high as (or higher than) $e^+e^-$ at LEP (which has no strongly
  interacting background) and $ep$ at HERA. This demonstrates that exclusive dileptons can be
  extracted from the huge backgrounds in hadron-hadron collisions, which is good news as they provide an excellent (probably the
  best) calibration of the momentum scale and resolution of the high precision proton spectrometers being planned~\cite{fp420} for
  ATLAS and CMS. One does not need to see both protons to calibrate the spectrometers,
  as each one is very well known (ultimately limited by the incoming beam momentum spread, $\frac{dp}{p}\sim10^{-4}$). The use of
  exclusive QED dileptons, with precisely known cross section, to measure the machine luminosity (integrated over a period, perhaps
  days), has been suggested. Unfortunately the precision is likely to be limited by unseen proton dissociation, and knowledge of
  efficiencies, acceptance and non-exclusive background. (One cannot require no pile-up, as the result
  would then depend on $\sigma_{inel}$ which is \emph{a priori} unknown).
  
     Exclusive $Z$ production is allowed in the Standard Model: a radiated virtual photon fluctuates to a $q\bar{q}$ pair,
     which scatters by hard pomeron exchange on the other proton, followed by $q\bar{q} \rightarrow Z$, as in exclusive
     vector meson photoproduction, see Fig. 1. However the SM cross section is much too small at the Tevatron: 
     0.3 fb~\cite{motyka} or 0.21
     fb~\cite{goncalvesz}. At the LHC (14 TeV) the predictions are 13 fb~\cite{motyka} and 69 fb~\cite{goncalvesz}, 
     which may make an observation
     possible. A signal at the Tevatron, or a significantly higher cross section at the LHC, would be evidence for new
     particles with strong and electroweak couplings. White's theory~\cite{white} of the supercritical pomeron predicts
     color-sextet quarks coupling strongly to the pomeron and to the $W$ and $Z$, and he expects a much enhanced cross
     section, but without a quantitative prediction. We used a sample of 3.17$\times 10^5$ lepton pairs with $M(l^+l^-) >$ 40 GeV/c$^2$,
     of which 1.83$\times 10^5$ were in the $Z$ peak. We required exclusivity over the full range $-7.4 < |\eta| < +7.4$,
     finding 8 events, agreeing with QED expectations. Fig. 2 shows the mass and azimuthal difference distributions of these
     events. All events were very back-to-back in the transverse plane, with $(\pi
     - \Delta\phi) < 0.75^\circ$. One event with $M(\mu^+\mu^-)$ = 66 GeV/c$^2$ had a $\bar{p}$ track in the Roman pots; for
     the others the $\bar{p}$ was out of their acceptance or they were not operational. None of the eight exclusive events were $Z$-candidates,
     and a limit was placed: $\sigma_{excl}(Z) <$ 0.96 pb at 95\% C.L.\footnote{Intriguingly, an 
     event with $M(e^+e^-)$ = 92 GeV/c$^2$, and $(\pi-\Delta\phi)= 1.25^\circ$ (larger than expected
     for QED) was rejected; it failed the exclusivity requirement only in the BSC counters on one arm. It has the
     characteristics of a
     photoproduced $Z$ but with a proton dissociation. However with only one event no claim can be made.} A nice check of the exclusivity analysis comes from
     $W\rightarrow l^\pm \nu$ events, which cannot be exclusive, but otherwise are very similar to $Z$, and which are more
     abundant. It may be possible to improve this limit using a factor 2-3 more data and including pile-up events, using the
     requirements of no associated tracks on the $l^+l^-$ vertex, small ($\pi - \Delta\phi$) and $p_T(l^+l^-)$. We are
     testing this method on the Upsilon region, with $8 < M(l^+l^-) < 40$ GeV/c$^2$. The QED is a good control, and the
     photoproduced $Y$ states have cross sections that are within reach, although not very well known. 
     (The HERA data~\cite{zeusy}
     do not resolve clearly the $Y(1S),Y(2S)$ and $Y(3S)$ states and have quite large uncertainties.)  
     Predictions for $\frac{d\sigma}{dy}(Y(1S),y=0)$ are~\cite{motyka,ykleinn,ybzdak,rybarska} are around 5 - 14 pb. Applying the
     branching fraction to $\mu^+\mu^-$ or $e^+e^-$, $B= 0.025$,
     would give a few hundred events in 2 fb$^{-1}$ ($\times$ the acceptance and
     efficiency). HERA has provided a nice compilation~\cite{herav} of exclusive cross sections for vector mesons from $\rho$ to
     $Y(1S)$ vs. $W = \sqrt{s(\gamma p)}$. In CDF with $y=0$ we have $W(J/\psi) \sim$ 80 GeV and $W(Y(1))\sim$ 136
     GeV. At HERA the ratio of these cross sections is $\sim$ 300. We are studying this
     region both in $\mu^+\mu^-$ and $e^+e^-$ events. 
     
     \begin{figure}[htp]
\centering
\subfigure[Mass spectrum of exclusive lepton pairs.]{
  \label{fig:qed}
  \includegraphics[width=.45\textwidth]{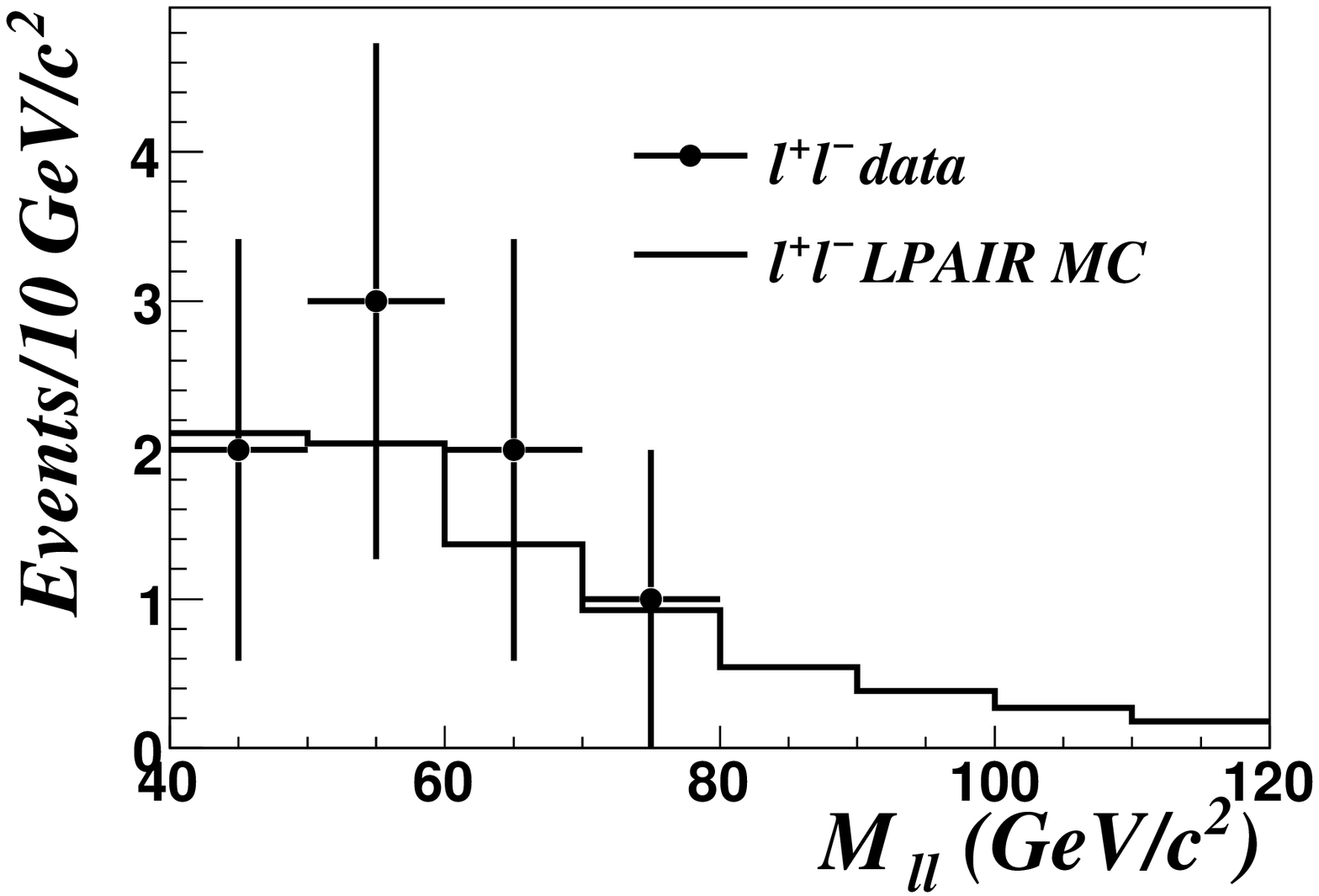}}
  \hspace{.3in}
  \subfigure[$180^\circ - \Delta\phi$ between the leptons.]{
  \label{fig:zphot}
  \includegraphics[width=.45\textwidth]{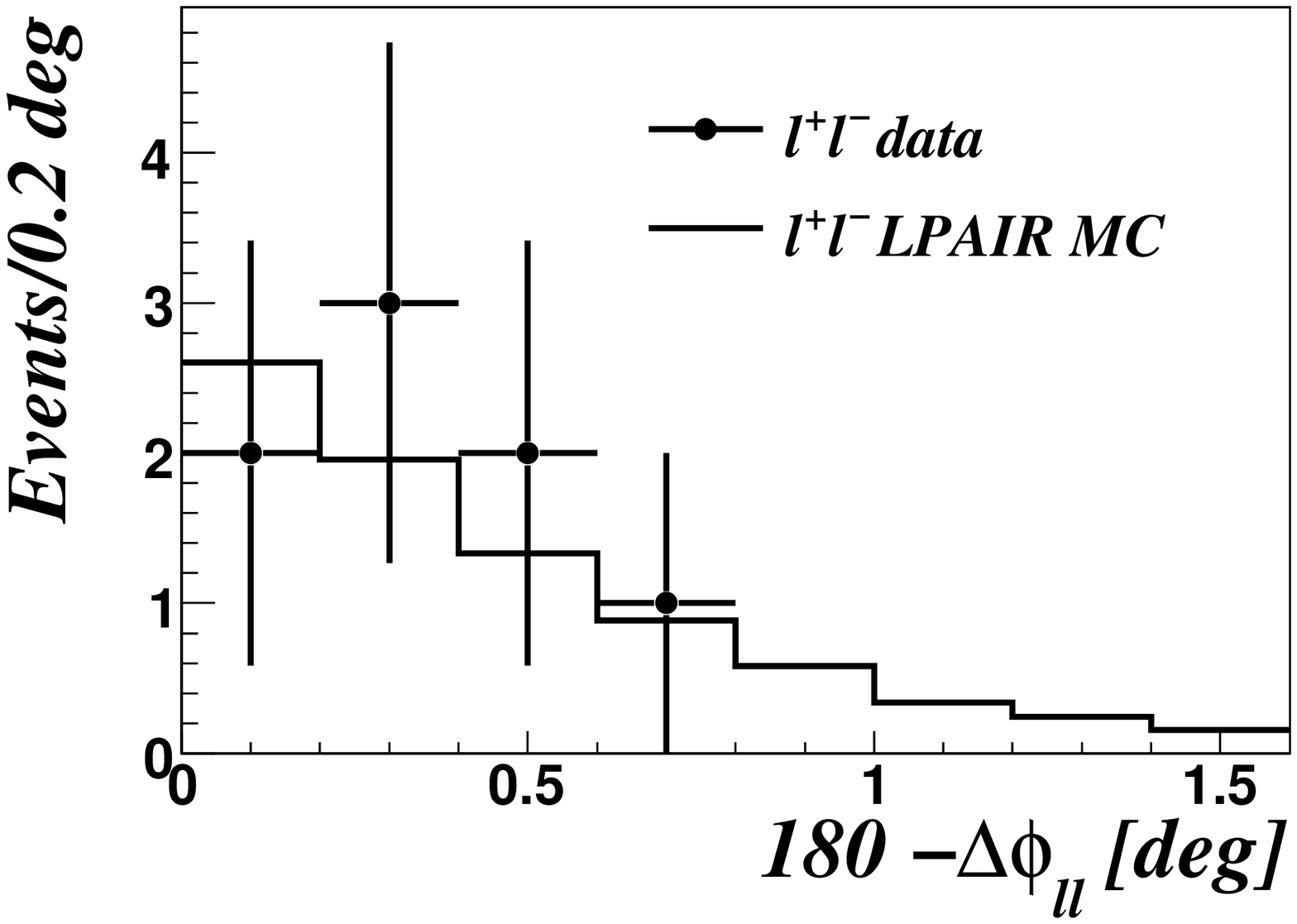}}
  \caption{Mass distribution of exclusive lepton pairs and azimuthal angle difference (from 180$^\circ$),
  data and normalized \textsc{lpair} prediction.}
  \end{figure}

     For the dimuons we used a trigger with two muons with $p_T(\mu) >$ 4 GeV/c and $|\eta| <$ 0.6. The inclusive
     $M(\mu^+\mu^-)$ spectrum shows the states $Y(1S),Y(2S)$ and $Y(3S)$ as well-separated peaks, see e.g. Ref.~\cite{upsinc};
     the mass resolution is only $\sigma(M) \approx$ 50 MeV/c$^2$, less than the mass differences. Separation is important
     in order to measure exclusive $\chi_b \rightarrow \Upsilon + \gamma$, which feed differently the three states.
     Inclusively the ratio $\Upsilon(1S)$:continuum is about 8:1.  
     Requiring no other tracks on the $\mu^+\mu^-$ vertex, $\pi - \Delta \phi <$ 0.1 rad
     and $p_T(\mu^+\mu^-) <$ 1.5 GeV/c retains both $\Upsilon$ and continuum events, but we cannot yet claim that these
     events are exclusive. These cuts should be efficient
     for the QED events and for most of the $Y$'s; the issue is non-exclusive backgrounds in the presence of other
     interactions. We are now studying the continuum to see if
     it is exactly as expected for QED, as a control, and can then give $Y$-photoproduction cross sections, depending
     on unknown backgrounds from $\chi_b \rightarrow Y + \gamma$. Unfortunately neither the $\chi_b$ production rates nor their
     radiative decays are known. The $p_T(Y)$ distribution is broader for $\chi_b$-daughters, which may help, but one
     would like to reconstruct the photons, which is probably not possible with pile-up. The same issues will confront
     us at the LHC.
     
     For the $Y \rightarrow e^+e^-$ channel we chose to veto pile-up, and used a trigger requiring two electromagnetic
     showers with $E_T > 2$ GeV and $|\eta| <$ 2m, and forward gaps (BSC-1 empty). Our prime motivation for this trigger is to search for
     additional exclusive $\gamma\gamma$ candidates, and hopefully to make a definitive observation. Compared with our earlier search, 
     we lowered the trigger threshold from 4 GeV to 2 GeV, took more data, will expand the $\eta$-coverage and use better
     background ($\pi^0$) rejection. Analysis is underway. The same trigger collects QED $e^+e^-$ pairs with $M(e^+e^-) >$ 8
     GeV/c$^2$, and $Y \rightarrow e^+e^-$ decays. The QED data is a good control of our exclusivity cuts, and the $Y$ candidates can be compared
     with our $Y \rightarrow \mu^+\mu^-$ candidates. As these events are without pile-up, we may have some $Y+\gamma$ candidates
     attributable to $\chi_b$ (but not many are expected). The decay photon tends to be soft (namely 391(442) MeV in the $Y$-frame) for the
     $\chi_{b0(b2)}$, and is poorly measured. Other exclusive final states are probably not useful, mainly because of trigger
     limitations.
     
     Can exclusive $Y$ photoproduction be seen with a forward proton tag at 420 m at the LHC? We 
     have $(1-x_F)_{1,2} = \frac{1}{\sqrt{s}}\sum_{leptons}
     p_T e^{+(-)\eta}$ ($1-x_F$ is the fractional momentum loss of the proton), and the most likely 
     kinematics are $p_{T,1} = p_{T,2} \approx 4.75$ GeV/c, $\eta_1 \approx \eta_2$. To have one proton with $1-x_F$ as large as (say) 0.01 we need a muon pair with $\eta \sim$ 2.5,
     at the limit of the muon coverage in CMS. A more serious problem with using $p+Y (\rightarrow \mu^+\mu^-) + p$ for 
     proton calibration is that many $\Upsilon$s, perhaps
     even most, will be decay products of $\chi_b$ states and the proton momenta are then not known. 
     For these reasons the QED $\gamma\gamma
     \rightarrow \mu^+\mu^-$ with $M(\mu^+\mu^-) \gtrsim  10$ GeV/c$^2$ will probably be the 
     calibration channel. Electron pairs are less favorable due to final state radiation and bremsstrahlung. With $|\eta_{max}| = 2$ and
     $M_{min}$ = 20 GeV/c$^2$ the cross section is 1.6 pb. One proton will 
     usually have much too small $\xi$, but fortunately both protons are
     known and one can be used to calibrate the spectrometers. 
     
     In conclusion, five years ago the predictions for exclusive Higgs production had more than two orders of magnitude spread. Since then in CDF
     we have measured three related processes, exclusive $JJ, \gamma\gamma$ and $\chi_c$, all consistent with the predictions of the
     Durham group within the quoted factor of ``a few". Especially the $\chi_c$ observation means that exclusive Higgs production
     must happen, if there is a Higgs boson. Our observation of exclusive photoproduced vector mesons demonstrates that exclusive
     $Z$ photoproduction must be possible, albeit with a small cross section (in the Standard Model) at the LHC. Our observations of exclusive
     $\gamma\gamma \rightarrow e^+e^-,\mu^+\mu^-$ (with a forward proton detected) are encouraging for forward spectrometer
     calibrations, and mean that $\gamma\gamma \rightarrow W^+W^-$ (and $\tilde{l^+}\tilde{l^-}$ if sleptons exist) could be measured. 
     This bodes well for a rich physics program with high precision forward spectrometers at the LHC.

\section{Acknowledgments}

I acknowledge support from the U.S. Dept.of Energy through Fermi National Accelerator Laboratory.

\end{document}